\documentclass[draftclsnofoot,12pt, onecolumn]{IEEEtran}
\usepackage{amsmath}
\usepackage{epsfig}
\usepackage{amssymb}
\usepackage{latexsym}
\usepackage{graphicx}
\usepackage{cite}

\title{On the $\kappa$-$\mu$/Gamma  Generalized Multipath/ Shadowing Fading Distribution}
\author{\IEEEauthorblockN{Paschalis C. Sofotasios\\}
\IEEEauthorblockA{School of Electronic and Electrical Engineering\\
University of Leeds, UK\\
Email: p.sofotasios@leeds.ac.uk\\}
\and
\IEEEauthorblockN{Steven Freear\\}
\IEEEauthorblockA{School of Electronic and Electrical Engineering\\
University of Leeds, UK\\
Email: s.freear@leeds.ac.uk}}
\begin{document}
\maketitle
\begin{abstract} 
This work is devoted to the formulation and derivation of the $\kappa-\mu$/gamma distribution which corresponds to A physical fading model. This distribution is composite and is based on the well known $\kappa-\mu$ generalized multipath model and the gamma shadowing model. A special case of the derived model constitutes the $\kappa-\mu$ Extreme/gamma model which accounts for severe multipath and shadowing effects. These models provide accurate characterisation of the simultaneous occurrence of multipath fading and shadowing effects. This is achieved thanks to the remarkable flexibility of their named parameters which render them capable of providing good fittings to experimental data associated with realistic communication scenarios. This is additionally justified by the fact that they include as special cases the widely known composite fading models such as Rice/gamma, Nakagami-m/gamma and Rayleigh/gamma. Novel analytic expressions are derived for the envelope and power probability density function of these distributions which are expressed in a relatively simple algebraic form which is convenient to handle both analytically and numerically. As a result, they can be meaningfully utilized in the derivation of numerous vital measures in investigations related to the analytic performance evaluation of digital communications over composite multipath/shadowing fading channels.
\end{abstract}

%
%
%
\section{Introduction}
\indent
It is widely known that fading is an effect that significantly degrades communication signals during wireless propagation. A common approach for accounting for this phenomenon has been viable through exploitation of appropriate statistical distributions. To this effect, statistical models such as Rayleigh, Nagakami-m, Weibull and Nakagami-q (Hoyt) have been shown to be capable of modelling small-scale fading in Non-Line-Of-Sight (NLOS) communication scenarios whereas Nakagami-$n$ (Rice) distribution has been typically utilized in characterizing multipath fading in Line-Of-Sight (LOS) communication scenarios, \cite{B:Nakagami, B:Jakes, B:Alouini} and the references therein. Capitalizing on the these models, M. D. Yacoub proposed three generalised fading models, namely, the $\alpha-\mu$, the $\kappa-\mu$, the $\eta-\mu$ models and subsequently the $\lambda-\mu$ and the $\kappa-\mu$ Extreme models, \cite{C:Yacoub_1, C:Yacoub_2, C:Yacoub_3, J:Yacoub_1, J:Yacoub_2, C:Yacoub_4, C:Rabelo}. These models have been proved particularly useful due to the remarkable flexibility offered by their named parameters which render them capable of providing adequate fittings to results obtained by field measurements. Their usefulness is also evident by the fact that they include as special cases all the above small-scale fading distributions. \\
\indent
It is recalled here that a fundamental principle of wireless radio propagation is that multipath and shadowing effects occur simultaneously. As a result, in spite of the undoubted usefulness of the aforementioned fading models, they all ultimately fail to account concurrently for both shadowing and multipath fading. In other words, the characterisation offered by the aforementioned fading models is limited to modelling either the one or the other effect. Based on this crucial limitation, the need for composite statistical models that can provide adequate characterization of the fading effect as a whole, became necessary \cite{B:Nakagami, B:Jakes, B:Alouini}. \\
\indent
Motivated by this, the authors in \cite{J:Kaveh} proposed the Rayleigh/gamma fading model, which is also known as $K$-distibution, $(K)$.  In the same context, Shankar in \cite{J:Shankar} exploited the flexibility of Nakagami-$m$ distribution, which includes Rayleigh distribution as a special case, and introduced the Nakagami-$m$/gamma composite distribution - or generalised $K$-distribution, $(K_{G})$. Likewise, the Weibull/gamma composite distribution was proposed in \cite{J:Bithas} while an introduction to generalised composite fading models  was reported by the authors in \cite{C:Sofotasios_1, C:Sofotasios_2, C:Sofotasios_3, Add, B:Sofotasios}. \\
\indent
The aim of this work is the derivation of novel analytic results for the $\kappa-\mu$/gamma and the $\kappa-\mu$ Extreme/gamma composite fading distributions. In more details, after formulating these models, novel analytic expressions are derived for their corresponding envelope and power probability density function (pdf).  The validity of the offered expressions is justified through comparisons with results obtained from numerical integrations while their behaviour is assessed under different parametric scenarios. Importantly, owing to the relatively simple algebraic representation of the offered expressions, they can be considered useful mathematical tools that can be efficiently utilized in studies related to the performance analysis of digital communications over $\kappa-\mu$/gamma and $\kappa-\mu$ Extreme/gamma composite fading channels. To this end, they can be exclusively employed in the derivation of analytic expressions for various performance measures such as error probability, channel capacity and higher order statistics, among others.\\
\indent
The remainder of this paper is organised as follows: Section II revisits the basic principles of the $\kappa-\mu$, the $\kappa-\mu$ Extreme and the gamma distributions. Subsequently, Sections III and IV are devoted to the formulation and derivation of the $\kappa-\mu$/gamma and $\kappa-\mu$ Extreme/gamma fading models, respectively, along with the necessary analysis on their behaviour. Finally, discussions on the potential applicability of the derived expressions in wireless communications along with closing remarks are given in Section V.  
 
%
%
\section{The $\kappa$-$\mu$, $\kappa$-$\mu$ Extreme and gamma Distributions}
\subsection{The $\kappa$-$\mu$ Fading Model}
The $\kappa-\mu$ distribution is a fading model that accounts for the small-scale variations of the fading signal in a line-of-sight (LOS) communication scenario and it is written in terms of two physical parameters, namely $\mu$ and $\kappa$. In terms of physical interpretation, the former parameter is related to the multipath clustering whereas the latter denotes the ratio between the total power of the dominant components and the total power of the scattered waves, \cite{C:Yacoub_2, J:Yacoub_2}. According to \cite{C:Yacoub_2}, for a fading signal with envelope $R$ and $\hat{r} = \sqrt{E(R^{2})}$, the $\kappa-\mu$ envelope pdf is given by,

\begin{equation} \label{1} 
p_{_{R}}(r) = \frac{2\mu (k+1)^{\frac{\mu + 1}{2}}}{k^{\frac{ \mu - 1}{2}}e^{\mu k}e^{\mu (1+k) \left(\frac{r}{\hat{r}} \right)^{2}}}\left(\frac{r}{\hat{r}}\right)^{\mu} 
I_{\mu - 1}\left(2\mu \sqrt{k(k+1)}\frac{r}{\hat{r}} \right)
\end{equation}
where $E(.)$ denotes expectation and $\hat{r}$ is the root-mean-square $(rms)$ value of $R$. Furthermore, the $I_{n}(x)$ function denotes the modified Bessel function of the first kind and order $n$ whereas the terms $\kappa >0$ and $\mu>0$ are obtained by \cite{C:Yacoub_2}

\begin{equation} \label{2} 
\mu = \frac{E^{2}(R^{2})}{Var(R^{2})}\frac{1 + 2k}{(1 + k)^{2}}
\end{equation}
which is valid for non-negative values. As already mentioned, the $\kappa{-}\mu$ distribution constitutes a particularly flexible multipath fading model since it includes as special cases the widely known Nakagami-n (Rice), Nakagami-m, Rayleigh and one-side Gaussian distributions, \cite{J:Yacoub_2}.\\
\subsection{The $\kappa$-$\mu$ Extreme Fading Model}
\indent
The $\kappa-\mu$ Extreme distribution emerges as a notable special case of the $\kappa-\mu$ distribution. Specifically, it is firstly recalled that all the aforementioned multipath models refer exclusively to conditions involving large numbers of paths which ultimately meet the principle of Central Limit Theorem (CLT). Nevertheless, it was shown recently that the occurrence of  large number of paths, is not typically the case in wireless radio propagation over enclosed environments such as aeroplanes, trains and buses \cite{B:Nakagami, J:Frolik, J:Durgin}. The reason underlying this is that such environments are known to be characterised by only a few number of paths. As a consequence, severe fading conditions, even worse than Rayleigh, are ultimately constituted \cite{C:Rabelo, J:Frolik, J:Durgin}. \\
\indent
By recalling \cite{J:Yacoub_1}, the Nakagami parameter $m$ is regarded as the inverse of the variance of the normalized power of the fading signal and is related to the parameters $\kappa$ and $\mu$ by the following relationship,

\begin{equation} \label{3} 
m = \frac{\mu (1+k)^{2}}{1+2k}
\end{equation}
At this point, the $\kappa-\mu$ Extreme distribution is obtained by keeping $m$ constant, assuming $\kappa \mu \approx 2m$ and considering a very strong LOS, i.e. $k\rightarrow \infty$, and very few multipaths, i.e. $\mu \rightarrow 0$. As a result, equation \eqref{1} reduces to \cite[eq. (16)]{C:Rabelo}, namely,

\begin{equation} \label{4} 
p_{_{R}}(\rho) = \frac{4mI_{1}(4m\rho)}{e^{2m(1+\rho^{2})}} + e^{-2m} \delta(\rho)
\end{equation}
where $p = r/\hat{r}$ while $\delta(.)$ denotes the Dirac delta function. \\
\subsection{The Gamma Fading Model}
\indent
It is recalled that log-normal distribution has been regarded the optimum statistical model for characterising the shadowing effect, \cite{B:Nakagami, B:Jakes, B:Alouini}. Nevertheless, in spite of its usefulness, it has been largely shown that when it is involved in combination with other elementary and/or special function, its algebraic representation becomes intractable. This is particularly the case in studies related to the analytical derivation of critical performance measures in the field of digital communications over fading channels. Motivated by this, the authors in \cite{J:Kaveh} proposed the gamma distribution as an accurate substitute to log-normal distribution. Mathematically, the envelope pdf of gamma distribution is given by \cite[eq. (4)]{J:Kaveh}, namely, 
 
\begin{equation} \label{5} 
p_{_{Y}}(y) = \frac{y^{b-1}e^{-\frac{y}{\Omega}}}{\Gamma(b) \Omega^{b}}, \qquad \, \, y\geq 0 
\end{equation}
where the term $b > 0$ is its shaping parameter and $\Omega = E(Y^{2})$. The gamma fading model has been shown to provide adequate fitting to experimental data that correspond to realistic fading conditions. In addition, it is evident that its algebraic representation is quite tractable and therefore, easy to handle both analytically and numerically. As a result, it has been undoubtedly useful in characterising shadowing and for this reason it has been exploited in the formulation of the $K$ and $K_{G}$ composite multipath/shadowing models, \cite{J:Kaveh, J:Shankar}.
%
%
\section{The $\kappa-\mu$/gamma Fading Distribution}
\subsection{Model Formulation}
\indent
According to the basic principles of statistics, the envelope pdf of a composite statistical distribution is deduced by means of superimpose of two or more statistical distributions. In the present case, this is realized by superimposing one multipath and one shadowing distribution. In mathematical terms,  this principle is expressed as,

\begin{equation} \label{6} 
p_{_{X}}(x) = \int_{0}^{\infty} p_{_{X\mid Y}}(x\mid y)p_{_{Y}}(y)dy
\end{equation}
where $p_{_{X\mid Y}}(x\mid y)$ denotes the corresponding multipath distribution with mode $y$. Based on this principle, the corresponding $\kappa-\mu$/gamma composite fading distribution is formulated by firstly setting $r = x$ and $\hat{r}=y$ in \eqref{1} and substituting  in \eqref{6} along with equation \eqref{5}. To this end, it follows immediately that

\begin{equation} \label{7} 
p_{_{X}}(x) = \frac{2\mu (1+k)^{\frac{\mu +1 }{2}}x^{\mu}}{\kappa^{\frac{\mu -1}{2}}e^{\mu \kappa}\Gamma(b)\Omega^{b}}\int_{0}^{\infty} \frac{I_{\mu - 1}\left(2\mu \sqrt{\kappa (1 + \kappa)} \frac{x}{y}\right)}{y^{1 + \mu -b}e^{\mu(1+\kappa)\frac{x^{2}}{y^{2}}}e^{\frac{y}{\Omega}}}dy
\end{equation}
Importantly, the term $y^{2}$ in \eqref{7} has emerged from the term $\hat{r}^{2}$ which denotes that the mean-squared value of the fading amplitude is gamma distributed. However,  it is noted here that it can be also assumed that the root-mean-squared value of the fading amplitude is gamma distributed. In fact, this is precisely the difference between the Rayleigh/Lognormal and Suzuki fading models since in the former the $rms$ value of the fading amplitude is modelled as being log-normal whereas in the latter it is the mean-squared value of this amplitude which is assumed to be log-normal \cite{B:Alouini}. As a result, by applying this principle in \eqref{7}, one obtains

\begin{equation} \label{8} 
p_{_{X}}(x) = \frac{2\mu (1+k)^{\frac{\mu +1 }{2}}x^{\mu}}{\kappa^{\frac{\mu -1}{2}}e^{\mu \kappa}\Gamma(b)\Omega^{b}}\int_{0}^{\infty} \frac{I_{\mu - 1}\left(2\mu \sqrt{\kappa (1 + \kappa)} \frac{x}{\sqrt{y}}\right)}{y^{1 + \frac{\mu}{2} -b}e^{\mu(1+\kappa)\frac{x^{2}}{y}}e^{\frac{y}{\Omega}}}dy
\end{equation}
\subsection{A Novel Expression for the Envelope pdf}
The derivation of an analytic expression for the envelope pdf of the $\kappa-\mu$ distribution is subject to evaluation of the integral in \eqref{7}. To this end, it is recalled the the $I_{\nu}(x)$ function can be equivalently represented in terms of the infinite series in \cite[eq. (8.445)]{B:Tables}, and the polynomial approximation in \cite[eq. (19)]{J:Gross}, namely,

\begin{align} \label{9} 
I_{\nu} (x) &= \sum_{l = 0}^{\infty} \frac{1}{\Gamma(l + 1) \Gamma(\nu + l + 1)} \left(\frac{x}{2} \right)^{\nu + 2l} \\
&\simeq \sum_{l = 0}^{n} \frac{\Gamma(n + l)}{\Gamma(l + 1) \Gamma(n - l + 1)}\frac{n^{1 - 2l}}{\Gamma(\nu + l + 1)} \left(\frac{x}{2} \right)^{\nu + 2l}
\end{align}
Therefore, by making the necessary change of variables\footnote{The polynomial approximation is used in the present analysis.}, one obtains straightforwardly\footnote{As $n \rightarrow \infty$, the polynomial approximation reduces to the infinite series.} 
 
\begin{equation} \label{10} 
I_{\mu - 1}\left(2\mu \sqrt{\kappa (1 + \kappa)}\frac{x}{\sqrt{y}} \right) \simeq  \sum_{l = 0}^{n}\frac{\Gamma(n + l)n^{1-2l}x^{\mu +2l -1}\mu^{\mu + 2l - 1}\kappa^{\frac{\mu -1}{2} + l}(1 + \kappa)^{\frac{\mu -1}{2}+l}}{\Gamma(l+1)\Gamma(n -l +1)\Gamma(\mu +l)y^{\frac{\mu -1}{2}+l}}
\end{equation}
Subsequently, by recalling that $\Gamma(x)\triangleq (x-1)!$ and substituting in \eqref{10} into \eqref{8}, it follows immediately that

$$
p_{_{X}}(x) =\sum_{l=0}^{n}\frac{2 \mu ^{\mu +2l} (1 + k)^{\mu + l}\kappa^{l} \Gamma(n +l)x^{2(\mu + l) -1}}{l!\Gamma(n -l +1)\Gamma(\mu + l)e^{\mu \kappa}n^{2l - 1}\Gamma(b) \Omega^{b}}
$$
\begin{equation} \label{11} 
\times \underbrace{\int_{0}^{\infty}y^{b - \mu - l - \frac{1}{2}} e^{-\mu(1 + \kappa)\frac{x^{2}}{y}}e^{-\frac{y}{\Omega}} dy}_{\mathcal{I}_{1}}
\end{equation}
Importantly, the above integral can be expressed in closed-form according to \cite[eq. (3.471.9)]{B:Tables}, namely,

\begin{equation} \label{12} 
\int_{0}^{\infty}x^{\nu - 1}e^{-\frac{b}{x}}e^{-\gamma x}dx = 2\left(\frac{b}{\gamma} \right)^{\frac{\nu}{2}}K_{\nu}\left(2\sqrt{b\gamma} \right)
\end{equation}
where $K_{\nu}(x)$ denotes the modified Bessel function of the second kind, \cite{B:Tables}. Therefore, by performing the necessary variable transformation in \eqref{12} and substituting in \eqref{11} yields the following explicit expression:

\begin{equation} \label{13} 
p_{_{X}}(x) = \left\lbrace \sum_{l=0}^{n}\frac{4\Gamma(n+l)n^{1 -2l}\mu^{\frac{\mu +3l +b}{2} +\frac{1}{4}} \kappa^{l}(1+\kappa)^{\frac{\mu +b +l}{2}+ \frac{1}{4}}}{l!\Gamma(n -l +1)\Gamma(\mu +l)e^{\mu \kappa} \Gamma(b)\Omega^{\frac{b +\mu +l	2}{2} - \frac{1}{4}}}  \, x^{\mu +l +b -\frac{1}{2}} K_{\mathcal{Z}}\left(2x\sqrt{\frac{\mu (1 +\kappa)}{\Omega}} \right) \right\rbrace + S\delta(x) 
\end{equation}
where $\delta(.)$ is the Dirac delta function and $\mathcal{Z} = b - \mu -l + \frac{1}{2}$. Furthermore, the notation $S$ denotes a normalisation scalar constant that needs to be determined so that \eqref{13} holds as a true pdf. To this end, by recalling that $\int_{0}^{\infty}p_{_{X}}(x) dx \triangleq 1$, it immediately follows that

\begin{equation} \label{14} 
S = 1 - \sum_{l=0}^{n}\frac{4\Gamma(n+l)n^{1 -2l}\mu^{\frac{\mu +3l +b}{2} +\frac{1}{4}} \kappa^{l}(1+\kappa)^{\frac{\mu +b +l}{2}+ \frac{1}{4}}}{l!\Gamma(n -l +1)\Gamma(\mu +l)e^{\mu \kappa} \Gamma(b)\Omega^{\frac{b +\mu +l	2}{2} - \frac{1}{4}}} \int_{0}^{\infty} x^{\mu +l +b -\frac{1}{2}} K_{\mathcal{Z}}\left(2x\sqrt{\frac{\mu (1 +\kappa)}{\Omega}} \right) dx
\end{equation}
Notably, the above integral can be expressed in closed-form according to \cite[eq. (6.561.16)]{B:Tables}. Based on this, the constant $S$ can be finally expressed as:

\begin{equation} \label{15} 
S = 1 - \sum_{l=0}^{n}\frac{\Gamma(n +l)\kappa^{l} \sqrt{\Omega}  n^{1 - 2l} \mu ^{l +\frac{1}{2}}\Gamma\left(\mu +\frac{1}{2} \right)}{l!\Gamma(n -l +1)e^{\mu \kappa}  \Gamma(b) 2^{1/4}} 
\end{equation}
Therefore, by substituting \eqref{15} into \eqref{13}, an explicit expression for the envelope pdf of the $\kappa-\mu$ fading distribution is deduced, namely,

\begin{equation} \label{16} 
p_{_{X}}(x) = 1 - \sum_{l=0}^{n}\frac{\Gamma(n +l) n^{1 -2l}\kappa^{l} e^{-\mu \kappa}}{l!\Gamma(n -l +1) \Gamma(b)} \times \left\lbrace \frac{\mu ^{\mu +\frac{1}{2}} \Gamma\left(\mu +\frac{1}{2} \right)}{2^{1/4} \Omega^{-1/2} } \frac{4 \mu ^{\frac{\mu +3l +b}{2} + \frac{1}{4}}x^{\mu +l +b -\frac{1}{2}}K_{\mathcal{Z}} \left(2x \sqrt{\frac{\mu (1 + \kappa)}{\Omega}} \right) }{\Gamma(\mu +l) (1 + \kappa)^{-\frac{\mu + b +l}{2} - \frac{1}{4}}\Omega^{\frac{b +\mu +l}{2} - \frac{1}{4}}} \right\rbrace
\end{equation}
which to the best of the authors' knowledge is novel.
\subsection{A Novel Expression for the Power pdf}
\indent
The corresponding power pdf of the $\kappa-\mu$ Distribution is straightforwardly deduced by making use of the relationship: $p_{_{W}}(w) = p_{_{P}}(\sqrt{w})/2\sqrt{w}$. To this effect, it follows that
 
\begin{equation} \label{17} 
p_{_{W}}(w) = 1 - \sum_{l=0}^{n}\frac{\Gamma(n +l) n^{1 -2l}\kappa^{l} e^{-\mu \kappa}}{l!\Gamma(n -l +1) \Gamma(b)} \times \left\lbrace \frac{\mu ^{\mu +\frac{1}{2}} \Gamma\left(\mu +\frac{1}{2} \right)}{2^{1/4} \Omega^{-1/2} } - \,\frac{2 \mu ^{\frac{\mu +3l +b}{2} + \frac{1}{4}}w^{\frac{\mu +l +b}{2} -\frac{3}{4}}K_{\mathcal{Z}} \left(2 \sqrt{\frac{\mu (1 + \kappa)}{\Omega}w} \right) }{\Gamma(\mu +l) (1 + \kappa)^{-\frac{\mu + b +l}{2} - \frac{1}{4}}\Omega^{\frac{b +\mu +l}{2} - \frac{1}{4}}} \right\rbrace
\end{equation}
\indent
Importantly, the algebraic form of the above expressions is relatively convenient to handle both analytically and numerically. This is particularly useful in deriving analytic expressions for various performance measures in telecommunications where the pdf of fading distributions is often involved in complex integrals combined with other elementary and/or special functions. 
%
%
\section{The $\kappa-\mu$ Extreme/gamma Fading Distribution}
\subsection{Model Formulation}
\indent
As already mentioned, the $\kappa-\mu$ Extreme fading model constitute a special case of the $\kappa-\mu$ distribution and it accounts for severe fading conditions that typically occur in enclosed environments. Likewise the $\kappa-\mu$ distribution, the $\kappa-\mu$ Extreme distribution does not account for the simultaneous occurrence of shadowing. Therefore, by following the same procedure as in the formulation of the $\kappa-\mu$/gamma distribution, the corresponding $\kappa-\mu$ Extreme/gamma multipath/shadowing composite fading model is formulated. To this end, equation \eqref{4} is equivalently expressed as:

\begin{equation} \label{18} 
p_{_{R}}(r) = 4m e^{-2m\left(1+\frac{r^{2}}{\hat{r}^{2}}\right)} I_{1}\left(4m \frac{r}{\hat{r}}\right) + e^{-2m} \delta(r)
\end{equation}
Next, by setting $r=x$ and considering that the (rms) of the the fading amplitude is gamma distributed, the $\hat{r}$ term in \eqref{18} is represented by $y$. Hence, by substituting on \eqref{6} along with \eqref{5}, it follows that the $\kappa-\mu$ Extreme/gamma distribution is given by:

\begin{equation} \label{19} 
p_{_{X}}(x) =  \frac{e^{-2m}}{\Gamma(b) \Omega^{b}} \left\lbrace \int_{0}^{\infty} \frac{4mI_{1}\left(4m\frac{x}{\sqrt{y}} \right)}{y^{1-b}e^{2m\frac{x^{2}}{y}} e^{\frac{y}{\Omega}}}dy +  \int_{0}^{\infty}\frac{e^{-\frac{y}{\Omega}}}{y^{1-b}}dy \right\rbrace
\end{equation}
\subsection{A Novel Expression for the Envelope pdf}
The second integral in \eqref{19} is expressed in closed-form in terms of the gamma function. As a result, it follows that

\begin{equation} \label{20} 
p_{_{X}}(x) = e^{-2m} + \frac{4me^{-2m}}{\Gamma(b)\Omega^{b}} \underbrace{\int_{0}^{\infty} \frac{e^{-2m\frac{x^{2}}{y}} e^{-\frac{y}{\Omega}}I_{1}\left(4m\frac{x}{\sqrt{y}} \right)}{y^{1-b}}dy}_{\mathcal{I}_{2}}
\end{equation}
The modified Bessel function in $\mathcal{I}_{2}$ can be re-written according to \cite[eq. (19)]{J:Gross}, namely, 

\begin{equation} \label{21} 
I_{1}\left(4m\frac{x}{\sqrt{y}} \right) = \sum_{l = 0}^{n}\frac{\Gamma(n +l)n^{1 -2l}2^{1 +2l}m^{1 +2l}x^{1 +2l}}{\Gamma(l+1)\Gamma(n - +1)\Gamma(l +2) y^{l +\frac{1}{2}}}
\end{equation}
 By substituting the above expression in yields,

\begin{equation} \label{22} 
p_{_{X}}(x) = e^{-2m}\left[ 1 + \sum_{l=0}^{n}\frac{m^{2(l+1)}\Gamma(n +l)n^{1 -2l}2^{2l +3}x^{1 +2l}}{l!\Gamma(n -l +1)\Gamma(l+2)\Gamma(b) \Omega^{b}} \right] \underbrace{\int_{0}^{\infty} y^{b - l - \frac{3}{2}} e^{-2m \frac{x^{2}}{y}}e^{-\frac{y}{\Omega}} dy}_{\mathcal{I}_{2}} \,\,+\,\, S
\end{equation}
where as in the case of the $\kappa-\mu$/gamma distribution, the scalar $S$ denotes a normalisation constant that constitutes the derived pdf true. \\
\indent
Notably, the $\mathcal{I}_{2}$ integral has the same algebraic form as the $\mathcal{I}_{1}$ integral in \eqref{11} and therefore in \eqref{12}, \cite[eq. (3.471.9)]{B:Tables}. To this effect, by making the necessary change of variables and substituting in \eqref{20} yields the following analytic expression: 
 
\begin{equation} \label{23} 
p_{_{X}}(x) =  e^{-2m}\left[1 + \sum_{l=0}^{n}\frac{\Gamma(n+l)n^{1 -2l}2^{\frac{b +3l}{2} + \frac{15}{4}}m^{\frac{b +3l}{2} + \frac{7}{4}}}{l!\Gamma(n -l +1)\Gamma(l+2)\Gamma(b)\Omega^{\frac{b +l}{2} + \frac{1}{4}}} \right]  x^{b +l +\frac{1}{2}} K_{ b -l - \frac{1}{2}}\left(2x\sqrt{\frac{2m}{\Omega}} \right) + S
\end{equation}
By recalling that $\int_{0}^{\infty} p(x)dx \triangleq 0$, it follows that 
 
\begin{equation} \label{24} 
S = 1 - e^{-2m}\left[1 + \sum_{l=0}^{n}\frac{\Gamma(n+l)n^{1 -2l}2^{\frac{b +3l}{2} + \frac{15}{4}}m^{\frac{b +3l}{2} + \frac{7}{4}}}{l!\Gamma(n -l +1)\Gamma(l+2)\Gamma(b)\Omega^{\frac{b +l}{2} + \frac{1}{4}}} \right]  \int_{0}^{\infty}x^{b +l +\frac{1}{2}} K_{ b -l - \frac{1}{2}}\left(2x\sqrt{\frac{2m}{\Omega}} \right)dx
\end{equation}
which with the aid of \cite[eq. (6.561.16)]{B:Tables}, yields
 
\begin{equation} \label{25} 
S = 1 - e^{-2m}  -e^{-2m} \sum_{l=0}^{n}\frac{\Gamma(n+l)n^{1 -2l}2^{\frac{l -b +3}{2}}m^{\frac{l -b}{2}+\frac{3}{4}}\Gamma\left(b + \frac{1}{2} \right)}{\Gamma(n- l +1)\Gamma(l +2)\Gamma(b)\Omega^{-\frac{b +l}{2} - \frac{5}{4}}} 
\end{equation}
Finally, by substituting \eqref{25} in \eqref{23}, an analytic expression for the envelope pdf of the $\kappa-\mu$ Extreme/gamma composite fading model, namely,

$$
p_{_{X}}(x) = 1 - e^{-2m} + \sum_{l=0}^{n} \frac{\Gamma(n +l)n^{1 -2l}e^{-2m}}{\Gamma(n -l +1)\Gamma(l +2)\Gamma(b)} \times
$$
\begin{equation} \label{26} 
\left\lbrace \frac{2^{\frac{b +3l}{2}+ \frac{15}{4}} K_{b -l -\frac{1}{2}}\left(2x\sqrt{\frac{2m}{\Omega}} \right) }{x^{-b -l -\frac{1}{2}} l!\Omega^{\frac{b +l}{2}+\frac{1}{4}}m^{-\frac{b +3l}{2}-\frac{7}{4}}}  - \frac{\Omega^{\frac{b+l}{2} +\frac{5}{4}} \Gamma\left(b +\frac{1}{2} \right)}{2^{\frac{b -l -3}{2}}m^{\frac{b -l}{2}- \frac{3}{4}}} \right\rbrace
\end{equation}
which to the best of the authors' knowledge, has not been previously reported in the open technical literature. 
\subsection{A Novel Expression for the Power pdf}
By recalling that $p_{_{W}}(w) = p_{_{P}}(\sqrt{w})/2\sqrt{w}$ and substituting accordingly in \eqref{26}, the corresponding power pdf of the $\kappa-mu$ Extreme/gamma fading model is deduced, namely,

\begin{equation} \label{27} 
p_{_{W}}(w) = \frac{1 - e^{-2m}}{2\sqrt{w}} + \sum_{l=0}^{n} \frac{\Gamma(n +l)n^{1 -2l}e^{-2m}}{\Gamma(n -l +1)\Gamma(l +2)\Gamma(b)} \left\lbrace \frac{2^{\frac{b +3l}{2}+ \frac{11}{4}} K_{b -l -\frac{1}{2}}\left(2\sqrt{\frac{2m}{\Omega}w} \right) }{w^{\frac{1}{4}-\frac{b +l}{2}} l!\Omega^{\frac{b +l}{2}+\frac{1}{4}}m^{-\frac{b +3l}{2}-\frac{7}{4}}}  - \frac{\Omega^{\frac{b+l}{2} +\frac{5}{4}} \Gamma\left(b +\frac{1}{2} \right)}{\sqrt{w} 2^{\frac{b -l -1}{2}}m^{\frac{b -l}{2}- \frac{3}{4}}} \right\rbrace
\end{equation}
%
%
\begin{figure}[h]
\centerline{\psfig{figure=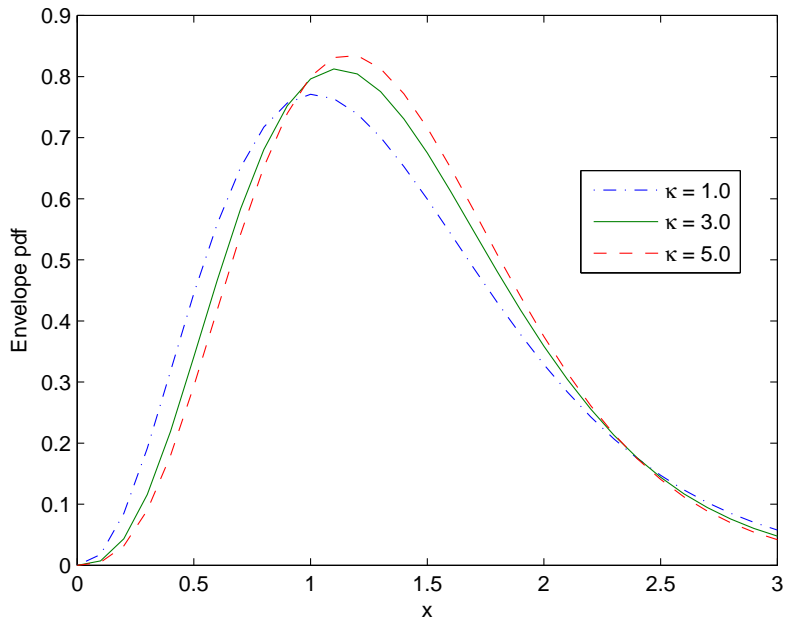,width=12.0cm, height=9.0cm}}
\caption{Envelope pdf of the $\kappa-\mu$/gamma distribution for $b=1.4$, $\Omega=1.2$, $\mu = 2.0$ and different values of $\kappa$}
\end{figure}

\section{Numerical Results and Discussions}

This Section is devoted to the demonstration of the general behaviour of the derived analytic expressions for the envelope pdf of the $\kappa-\mu$/gamma and $\kappa-\mu$ Extreme/ gamma fading distributions. To this end, Figure $1$ illustrates the pdf of $\kappa-\mu$/gamma with respect to $x$ for $b=1.4$, $\Omega=1.2$, $\mu = 2.0$ and different values of the parameter $\kappa$. Likewise, Figure $2$ considers $b=1.4$, $\Omega=1.2$, $\kappa = 1.0$ and different values of $\mu$. In the same context, Figure $3$ demonstrates the envelope pdf of the $\kappa-\mu$ Extreme /gamma distribution for $b=1.2$, $\Omega=0.8$ and different values of the Nakagami parameter $m$. One can observe the flexibility of the proposed models which, as already mentioned, render them capable of providing adequate fittings to experimental results. 
\begin{figure}[h]
\centerline{\psfig{figure=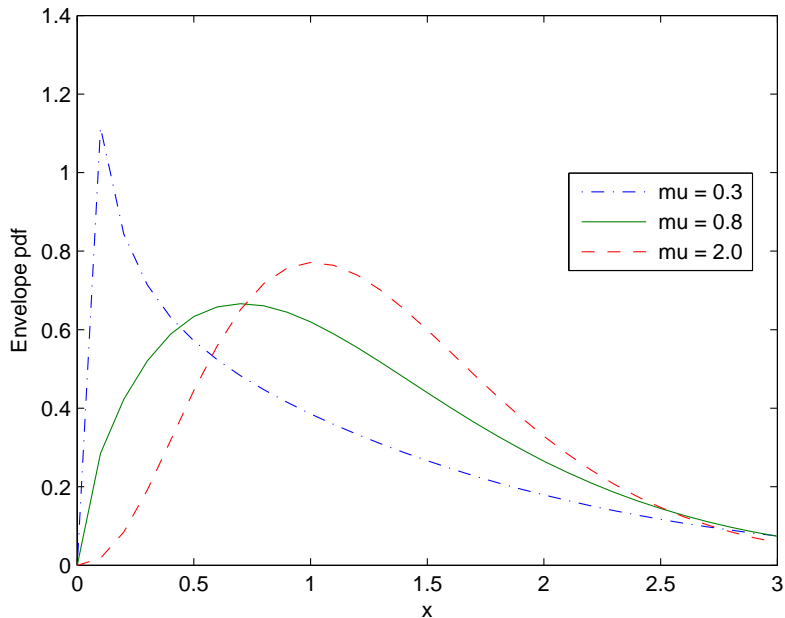,width=12.0cm, height=9.0cm}}
\caption{Envelope pdf of the $\kappa-\mu$/gamma distribution for $b=1.4$, $\Omega=1.2$, $\kappa = 1.0$ and different values of $\mu$}
\end{figure}
\begin{figure}[h]
\centerline{\psfig{figure=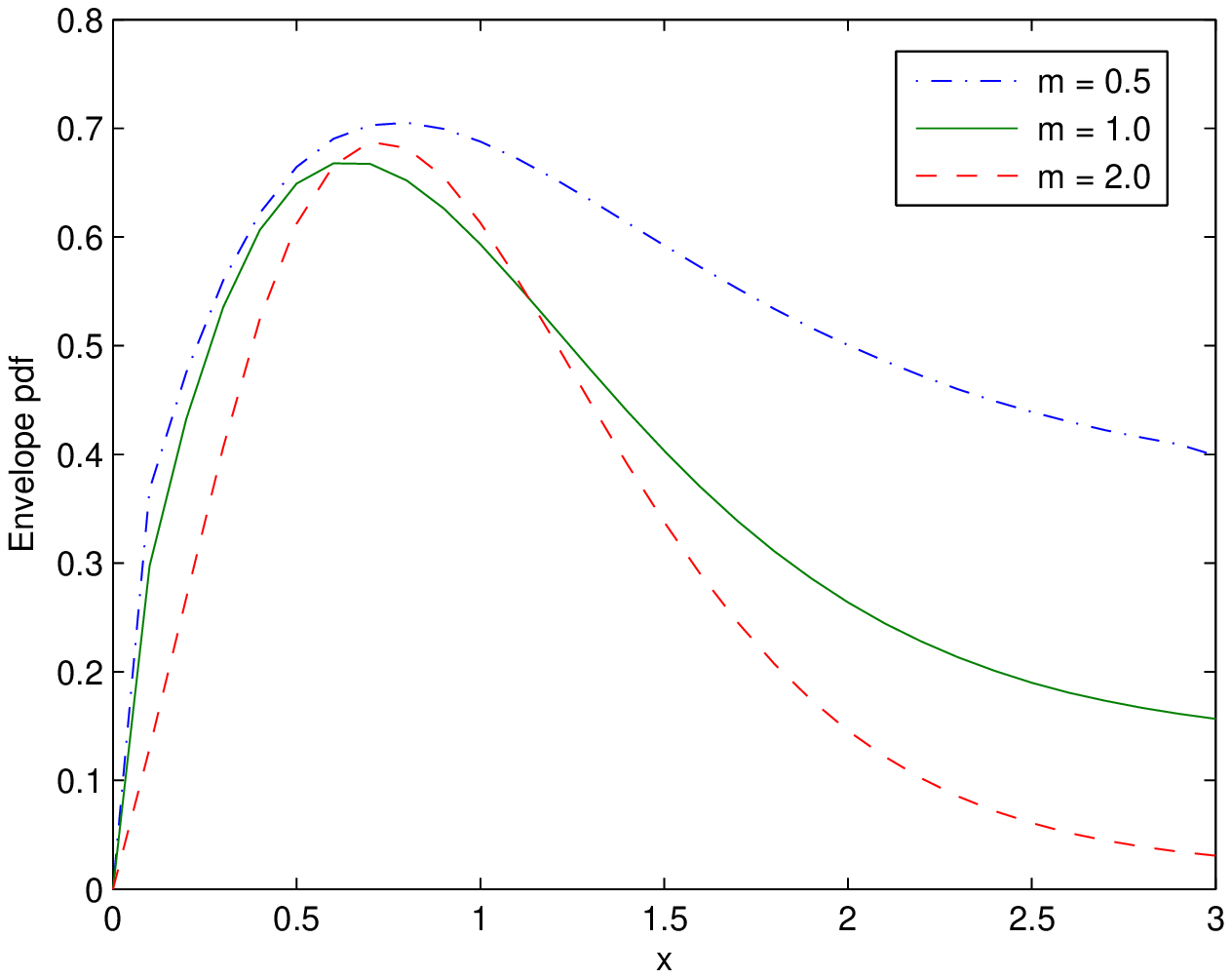,width=12.0cm, height=9.0cm}}
\caption{Envelope pdf of the $\kappa-\mu$ Extreme/gamma distribution for $b=1.2$, $\Omega=0.8$ and different values of $m$}
\end{figure}
\subsection{Usefulness and Applicability in Wireless Communications}
It is widely known that the algebraic representation of crucial performance measures is critical in studies related to analytical performance evaluation of digital communications. This is obvious by the fact that when the algebraic form of a corresponding measure is convenient, then it is ultimately more possible that the derived relationships be expressed in closed-form. Based on this, the fact that the form of the offered analytic expressions is relatively simple, renders the proposed fading models convenient to handle both analytically and numerically. Therefore, the derived expressions can be efficiently applied in various analytic studies relating to the performance evaluation of digital communications over composite multipath/shadowing fading channels including severe fading conditions. Indicatively, they can be meaningfully utilized in deriving explicit expressions for important performance measures such as, error probability, probability of outage, ergodic capacity, channel capacity under different constraints and higher order statistics. It is recalled here that expressions corresponding to the aforementioned measures can obviously be derived in both classical and emerging technologies such as single channel and multichannel communications, cognitive radio and cooperative systems and optical communications.
\section{Closing Remarks}
$ $\\
\indent
This work was devoted in the introduction, formulation and derivation of the $\eta-\mu$/gamma and $\lambda-\mu$/gamma fading distributions. These models are formulated from the superimpose of the $\eta-\mu$ and $\lambda-\mu$ generalised small-scale fading models, respectively, and the gamma shadowing model. These distribution are particularly flexible and it that includes as special cases the  widely known Hoyt, Nakagami-$m$ and Rayleigh fading models. Novel analytic expressions were derived for the envelope probability density function which can be considered a useful mathematical tool in applications related analytical performance evaluation of digital communications over multipath/shadowing channels. 
\bibliographystyle{IEEEtran}

\thebibliography{99}
%
\bibitem{B:Nakagami}  
M. Nakagami, 
\emph{The m-Distribution - A General Formula of Intensity Distribution of Rapid Fading}, W. C. Holfman, Ed. Statistical Methods in Radio Wave Propagation, Elmsford, NY, Pergamon, 1960.
\bibitem{B:Jakes}  
W. C. Jakes, 
\emph{Microwave Mobile Communications.} IEEE Computer Society Press, 1994.
\bibitem{B:Alouini} 
M. K. Simon, M-S. Alouini,
\emph{Digital Communication over Fading Channels}, New York: Wiley, 2005
\bibitem{C:Yacoub_1} 
M. D. Yacoub,
\emph{The $\eta$-$\mu$ Distribution: A General Fading Distribution}, IEEE Boston Fall Vehicular Technology Conference 2000, Boston, USA, Sep. 2000
\bibitem{C:Yacoub_2} 
M. D. Yacoub,
\emph{The $\kappa$-$\mu$ Distribution: A General Fading Distribution}, IEEE Atlantic City Fall Vehicular Technology Conference 2001, Atlantic City, USA, Oct. 2001
\bibitem{C:Yacoub_3} 
M. D. Yacoub,
\emph{The $\alpha$-$\mu$ distribution: A General Fading Distribution}, in Proc. IEEE Int. Symp. PIMRC, Sep. 2002, vol. 2, pp. 629–633.
\bibitem{J:Yacoub_1} 
M. D. Yacoub,
\emph{The $\alpha$-$\mu$ Distribution: A Physical Fading Model for the Stacy Distribution}, in IEEE Trans. Veh. Tech., vol. 56, no. 1, Jan.2007
\bibitem{J:Yacoub_2} 
M. D. Yacoub,
\emph{The $\kappa$-$\mu$ distribution and the $\eta$-$\mu$ distribution}, in IEEE Antennas and Propagation Magazine, vol. 49, no. 1, Feb. 2007.
\bibitem{C:Yacoub_4} 
G. Fraidenraich and M. D. Yacoub,
\emph{The $\lambda-\mu$ General Fading Distribution}, in Proc. IEEE Microwave and Optoelectronics Conference. IMOC 2003. Proceedings of the  SBMO/IEEE - MTT-S International, pp. 249-254, 2003 
\bibitem{C:Rabelo} 
G. S. Rabelo, U. S. Dias, M. D. Yacoub,
\emph{The $\kappa$-$\mu$ Extreme distribution: Characterizing severe fading conditions}, in Proc. SBMO/IEEE MTT-S IMOC, 2009
\bibitem{J:Kaveh} 
A. Abdi and M. Kaveh,
\emph{K distribution: an appropriate substitute for Rayleigh-lognormal distribution in fading shadowing wireless channels}, Elec. Letters, Vol. 34, No. 9, Apr. 1998
\bibitem{J:Shankar} 
P. M. Shankar,
\emph{Error rates in generalized shadowed fading channels}, Wireless Personal Communications, vol. 28, no. 4, pp. 233–238, Feb. 2004
\bibitem{J:Bithas} 
P. S. Bithas
\emph{Weibull-gamma composite distribution: Alternative multipath/shadowing fading model}, in Electronic Letters, Vol. 45, No. 14, Jul. 2009
\bibitem{C:Sofotasios_1} 
P. C. Sofotasios, S. Freear,
\emph{The $\kappa$-$\mu$/gamma composite fading model}, in Proc. IEEE ICWITS, Aug. 2010, Hawaii, USA
\bibitem{C:Sofotasios_2} 
P. C. Sofotasios, S. Freear,
\emph{The $\eta$-$\mu$/gamma composite fading model}, in Proc. IEEE ICWITS, Aug 2010, Hawaii, USA
\bibitem{C:Sofotasios_3} 
P. C. Sofotasios, S. Freear,
\emph{The $\kappa$-$\mu$ Extreme/Gamma Distribution: A Physical Composite Fading Model}, in IEEE WCNC 2011, Cancun, Mexico, pp. 1398 - 1401, March 2011

 \bibitem{Add}
S. Harput, P. C. Sofotasios, S. Freear, 
``Novel Composite Statistical Model For Ultrasound Applications,''
\emph{IEEE IUS '11}, pp. 1387$-$1390, Orlando, FL, USA, Oct. 2011. 

\bibitem{B:Sofotasios} 
P. C. Sofotasios,
\emph{On Special Functions and Composite Statistical Distributions and Their Applications in Digital Communications over Fading Channels}, Ph.D Dissertation, University of Leeds, UK, 2010
\bibitem{J:Frolik} 
L. Bakir and J. Frolik, 
\emph{Diversity gains in two-ray fading channels}, IEEE Trans. Wirel. Commun., vol. 8, no. 2, pp. 968-977,2009.
\bibitem{J:Durgin} 
G. Durgin, T. Rappaport, and D. de Wolf, 
\emph{New analytical models and probability density functions for fading in wireless communications}, In IEEE Trans Commun., vol. 50, no. 6, pp. 1005-1015,
2002
\bibitem{B:Tables} 
I. S. Gradshteyn and I. M. Ryzhik, 
\emph{Table of Integrals, Series, and Products}, $7^{th}$ ed. New York: Academic, 2007.
\bibitem{J:Gross} 
L- L. Li, F. Li and F. B. Gross,
\emph{A new polynomial approximation for $J_{m}$ Bessel functions}, Elsevier journal of Applied Mathematics and Computation, Vol. 183, pp. 1220-1225, 2006

\end{document}